\documentclass[a4paper]{jpconf}
\usepackage{graphicx}

\def\beq{\begin{equation}} \def\eeq{\end{equation}}
\def\bea{\begin{eqnarray}} \def\eea{\end{eqnarray}}

\begin{document}
\title{
Observables in classical canonical gravity: folklore demystified}

\author{J. M. Pons}

\address{Departament d'Estructura i Constituents de la Mat\`eria and Institut de
Ci\`encies del Cosmos, Universitat de Barcelona,
Diagonal 647, 08028 Barcelona, Catalonia, Spain}

\ead{pons@ecm.ub.es}
\author{D. C. Salisbury}

\address{Department of Physics,
Austin College, Sherman, Texas 75090-4440, USA\\
Max-Planck-Institut f\"ur Wissenschaftsgeschichte,
Boltzmannstrasse 22,
14195 Berlin, Germany}

\ead{dsalisbury@austincollege.edu}
\author{K. A. Sundermeyer}

\address{Freie Universit\"at Berlin, Fachbereich Physik,
Institute for Theoretical Physics, Arnimallee 14, 14195 Berlin,
Germany}

\ead{ksun@gmx.de}

\begin{abstract}
We give an overview of some conceptual difficulties, sometimes called paradoxes, that have puzzled for years the physical interpetation of classical canonical gravity and, by extension, the canonical formulation of generally covariant theories. We identify these difficulties as stemming form some terminological misunderstandings as to what is meant by ``gauge invariance", or what is understood classically by a ``physical state". We make a thorough analysis of the issue and show that all purported paradoxes disappear when the right terminology is in place. Since this issue is connected with the search of observables - gauge invariant quantities - for these theories, we formally show that time evolving observables can be constructed for every observer. This construction relies on the fixation of the gauge freedom of diffeomorphism invariance by means of a scalar coordinatization. We stress the condition that the coordinatization must be made with scalars. As an example of our method for obtaining observables we discuss the case of the massive particle in AdS spacetime.
\end{abstract}

%\pacs{4.20.Fy, 4.60.Ds.}

\section{Introduction}

A menacing paradox 
has been haunting for years the canonical formulation of generally covariant theories. Briefly, the paradox is the following: since the Hamiltonian is nothing but a particular case of the gauge 
generator, one can - or must - be led to conclude that there is no physical time evolution in such  
theories, hence ``time is frozen". A way of rephrasing this  paradox is to argue that the  
observables, being by definition gauge invariant, must be constants of motion, hence ``nothing happens".

So the question is: is there something wrong in the canonical formulation of generally covariant theories?  
Our answer - which we believe is {\it the} answer - is: certainly not!. In 
\cite{Pons:2009cd,Pons:2009cz}, building on previous work in \cite{ps04}, we have shown that observables 
evolving in time can be constructed\footnote{The construction is formal, but solves the conceptual issue. 
See the conclusions for more comments on this point.}. We have argued that there is nothing wrong with 
the canonical formulation, and we have located the origin of the problems in some misunderstandings, basically 
terminological, that have over the course of time morphed into conceptual ones. We do not want to repeat verbatim here the arguments spelled out in the previous publications, but we will rather try to give an overview of the subject, emphasizing the conceptual aspects as well as other issues not covered in these publications, and refer to \cite{Pons:2009cz} for technical details when necessary. The - new - example of the massive particle in an anti-DeSitter (AdS) background will serve to illustrate our results. 

We begin in section 2 with some very general aspects of the canonical formulation for general relativity
theories and its gauge group (generally covariant theories can be similarly dealt with). In section 3 we 
 introduce and address the purported paradoxes sometimes claimed to plague the formalism. In 
section 4 we consider the coordinate dependent gauge fixings that will pave the way to the construction of 
observables given in section 5. Section 6 is devoted to the example, and section 7 to the conclusions.

\section{Diffeomorphism in canonical gravity}
%%%%%%%%%%%%%%%%%%%%%%%%%%%%%%%%%%%%%%%%%%%%%%%%%%%%%%%%%%%%%%
%%%%%%%%%%%%%%%%%%%%%%%%%%%%%%%%%%%%%%%%%%%%%%%%%%%%%%%%%%%%%%

It was clarified long ago \cite{KB} that the gauge group of general relativity (GR) - and of other 
generally covariant theories - is much larger than the diffeomorphism group. In fact, the scalar density 
character of the Lagrangian ${\cal L}$ ensures the Noether symmetry condition 
\beq
\delta {\cal L} = 
\partial_\mu(\epsilon^\mu {\cal L})
\label{gencov} 
\eeq 
for active variations of the fields induced by any {\sl field-dependent} infinitesimal coordinate 
transformation $x^\mu\to x^\mu - \epsilon^\mu(x,\phi(x))$, with $\phi$ representing any field or field 
component. We call this gauge group the diffeomorphism-induced gauge group. Theories satifying 
(\ref{gencov}) are commonly called generally covariant or diffeomorphism invariant. Of course the 
diffeomorphism group, with $\epsilon^\mu$ depending on the coordinates only, is just a subgroup of the 
gauge group.

It was also clarified in \cite{KB} that the diffeomorphism group can not be completely realized in phase 
space. What can be realized is a proper subgroup of the diffeomorphism-induced gauge group - which is 
sometimes called the Bergmann-Komar group, as we denote it here - in which the field dependent infinitesimal 
diffeomorphisms $\epsilon^\mu$ are of the specific form 
\beq 
\epsilon^\mu(x,\phi(x)) = n^\mu(x) \xi^0 + \delta^\mu_a \xi^a  
\label{rule} 
\eeq 
with $n^\mu =(N^{-1}, -N^{-1} N^a )$, where $N$ is the lapse and $N^a$ the shift, and $\xi^\mu$ are 
arbitrary infinitesimal functions of the coordinates as well as of the fields other than the lapse and 
shift. The argument given in \cite{KB} for this formula (\ref{rule}), see also \cite{Salisbury:1982ez}, 
relies on the requirement of consistency between the algebra of diffeomorphisms and the Poisson bracket 
structure in phase space. It was realized later \cite{pss1997pr} that an argument of projectability from 
configuration-velocity (or tangent space) structures to phase space leads to the same result in a simpler 
way. 

Note that although the gauge group of canonical gravity, the Bergmann-Komar group, is not as large as the 
gauge group in configuration-velocity space, every infinitesimal diffeomorphism between two given 
configurations in tangent space can be matched in phase space by an element belonging to the Bergmann-Komar 
group. In this sense the Bergmann-Komar group is large enough in order to to fully describe general 
covariance in phase space. 

Using the Rosenfeld-Dirac-Bergmann formalism\footnote{See \cite{Salisbury:2007br} and also 
\cite{Salisbury:2010} in these proceedings for an evaluation of the 
long forgotten Rosenfled's contribution.} for gauge theories in phase space, the Dirac Hamiltonian of 
general relativity becomes \cite{ADM} (here and henceforth sum of repeated indices includes spatial 
integration as well)
\beq 
   H_{\!{}_D} =   N^\mu {\cal H}_{\mu} + \lambda^\mu P_{\mu}\,,
\label{theham}
\eeq
where $P_{\nu}$ are the canonical momenta of the lapse $N=N^0$ and shift 
$N^i$, ${\cal H}_{\mu}$ are the Hamiltonian and momentum constraints, and $\lambda^\mu$ are arbitrary 
functions. The dynamics imposes $\dot N^\mu = \lambda^\mu$ and thus one deduces that the lapse and shift are  basically arbitrary functions. This is the reason why one usually assigns them values as part of a gauge fixing; for 
instance one can take the popular $N=1$ and $N^i=0$. This partial gauge fixing results in cutting down the 
Bergmann-Komar gauge group and in particular the infinitesimal gauge generators. The common use of  
partially gauge fixed actions in the phase space formulation explains why it took so long since the 
pioneering work of \cite{KB} until the entire gauge generator of the Bergmann-Komar group was 
formulated. It takes the form \cite{pss1997pr} 
\beq 
G_{{\xi }}(t) =  ( {\cal H}_{\mu}
+ N^{\rho} C^{\nu}_{\mu \rho} P_{\nu}) \xi^{\mu} +  P_{\mu} \dot\xi^{\mu} \,,
\label{thegen}
\eeq
(the dot stands for time derivative) where $\xi^{\mu}$ are the arbitrary functions - henceforth called 
descriptors - associated with an infinitesimal spacetime diffeomorphism through (\ref{rule}), and 
$C^{\nu}_{\mu \rho}$ are the structure functions resulting form the algebra of the Hamiltonian and momentum 
constraints under the Poisson bracket. Note that (\ref{thegen}) is the generator of gauge transformations 
for all phase space variables, including the lapse and shift. One can then explicitely check (\ref{rule}) 
for all the variables. The Poisson brackets of $G_{{\xi }}$ with different descriptors exhibit a soft 
algebra structure, with structure functions instead of structure constants. The reason can be traced 
\cite{Teitel,Salisbury:1982ez,Pons:2003uc} to the relations (\ref{rule}). 

This gauge generator $G_{{\xi }}$ realizes infinitesimally the full active diffeomorphism invariance of general relativity (GR) 
in phase space (for diffeomorphisms connected with the identity). It has the four arbitrary functions - in 
four dimensions - corresponding to four-diffeomorphism invariance. A good gauge fixing - more on this later 
- would yield the standard counting of degrees of freedom of GR. In conclusion, as regards the gauge 
symmetries, the canonical formalism is equivalent to the tangent space formalism, with the caveat 
concerning the aforementioned projectability issue.

\section{A debate in canonical gravity}
%%%%%%%%%%%%%%%%%%%%%%%%%%%%%%%%%%%%%%%%%%%%%%%%%%%%%%%%%%%%%%
%%%%%%%%%%%%%%%%%%%%%%%%%%%%%%%%%%%%%%%%%%%%%%%%%%%%%%%%%%%%%%

The classical canonical formalism of general GR and, by extension, of any generally covariant 
theory, has been the subject of a long debate regarding its physical intepretation. This debate has not 
only affected the physicists practitioners in the field but also the philosophers' community, see for 
instance \cite{earman,maudlin} and references therein. Let us briefly present the two sides of the debate. 

The position on one side is that there ought to be no debate at all because the phase 
space formalism is equivalent to the formalism in configuration-velocity space, and no one has claimed that 
any interpretational problem exists in the latter framework. Entire books have been devoted to the 
experimental tests of GR, and this very language implies that observables exist - alive and kicking. Thus 
the entire debate must be a consequence of misunderstandings.

On the other side of this debate it has been claimed that the phase space formalism suffers from some 
difficulties regarding the physical intepretation of what constitutes an observable in these theories. Here  we tend to be confronted with paradoxical slogans like ``frozen time"  or the equally mysterious ``nothing 
happens" picture. According to this view there can no be real physical time evolution for these theories. In 
another contribution to these proceedings \cite{Salisbury:2010} it is shown that this problem already long ago attracted the interest of eminent physicists like Peter Bergmann and P.A.M. Dirac. Without 
mathematical details, which will be given later, let us briefly recall here  the two basic 
arguments that are wielded on this side of the debate.

\subsection{The issue of {\sl gauge versus dynamics}}
\label{gvd}
%%%%%%%%%%%%%%%%%%%%%%%%%%%%%%%%%%%%%%%%%%%%%%%%%%%%%%%%%%%%%%
%%%%%%%%%%%%%%%%%%%%%%%%%%%%%%%%%%%%%%%%%%%%%%%%%%%%%%%%%%%%%%

The gauge generator, which depends on some arbitrary functions  (the descriptors), generates active diffeomorphism-induced variations of phase space variables.But this generator becomes exactly the Dirac Hamiltonian when 
the descriptors take some specific vaules, actually the values of the lapse and shift. Then the argument 
goes: if the Dirac Hamiltonian is just a particular case of the gauge generator, this means that the time 
evolution is just gauge, hence ``nothing happens". ``Time is frozen". (Incidentally,  
boundary terms in the Dirac Hamiltonian, or in the gauge generators, leave this puzzling situation 
unaffected, because such terms have no effect either on the gauge transformations or on the dynamics.) 

A variant of this argument proceeds by saying that if one quotients out the gauge-equivalent points in phase 
space so as to eliminate the gauge freedom, points that represent future - or past - evolution with 
respect to some initial point, belong to the same equivalence class and are thus identified. Hence 
``time is frozen".

\subsection{The issue of {\sl observables as constants of motion}}
\label{oct}

Obervables in phase space must be functionals of the fields having vanishing Poisson brackets with the 
gauge generator. Consequently, as we have seen, they must have vanishing Poisson brackets with the 
Dirac Hamiltonian as well; therefore the observables can not have implicit time dependence. On the 
other hand, because of general covariance, gauge-invariant quantities must be independent of the 
coordinates and thus in particular they can not have explicit time dependence. Both considerations together 
imply that the observables must be constants of motion without explicit dependence on the coordinates. 
Hence we encounter again the paradoxical ``nothing happens".

\vspace{6mm}

Taken at face value, the case for the ``nothing happens" picture seems quite strong, perhaps even
insurmountable. We argue that this picture arises in fact from some regrettable
misunderstandings, particularly on the use of some basic terms like ``gauge invariance", or even the more 
basic notions of  ``gauge symmetry" or of what is understood classically by a ``physical state". Once we 
agree on the terminology, clarifications immediately follow and the paradoxical picture quietly fades away. 
The case is closed.

Now we will address and resolve the two paradoxes just mentioned. 

\subsection{addressing the issue of {\sl gauge versus dynamics}}
%%%%%%%%%%%%%%%%%%%%%%%%%%%%%%%%%%%%%%%%%%%%%%%%%%%%%%%%%%%%%%
%%%%%%%%%%%%%%%%%%%%%%%%%%%%%%%%%%%%%%%%%%%%%%%%%%%%%%%%%%%%%%

The teminology we propose is the standard one. A symmetry is a map sending solutions (entire on-shell field 
configurations) of the equations of motion (EOM) to solutions. The symmetry will be gauge if it depends on 
arbitrary functions of the spacetime coordinates and the fields as well. This terminology is standard, as 
we say, but let us immediately warn the reader that an important source of confusion can be traced to the 
influential book by Dirac \cite{Dira64}, in which gauge tranformations at a fixed time $t_0$ are 
considered. Of course there is nothing wrong with it as an alternative definition, but it is a concept rather different from that of 
mapping solutions to solutions \cite{Pons:2004pp}. 

Since gauge tranformations map solutions to solutions, our natural arena will be the space ${\cal S}$ of 
on-shell field configurations, i.e., fields obeying the EOM - which we take in phase space.  This space is 
a subset of the much bigger space of general field configurations. An infinitesimal gauge transfomation 
acts on this bigger space with the ordinary Poisson bracket\footnote{Note that the Poisson bracket is an 
equal time action. One way of defining its action is to employ a one-parameter family of phase space variables, with parameter $t$.}, and its action can be restricted to ${\cal S}$ because the generators 
of gauge transformations define an action which is tangent to ${\cal S}$. A point $p$ in ${\cal S}$ is an 
entire spacetime\footnote{Spacetimes will always be understood as  encompassing also their on-shell  field
configurations.} with the fields - solution of the EOM - described in a particular coordinatization. For 
practical purposes, though, it will be enough to work in a coordinate patch of a given chart. To every 
point $p$ there is associated an ``observer", or ``user", who is using such a coordinatization to describe 
the fields in spacetime.

The gauge generator, acting through the equal time Poisson bracket, produces a symmetry - mapping solutions 
to solutions - when all times $t$ are considered in (\ref{thegen}). We need $G_{{\xi }}(t)$, for all times $t$, to 
move from a point $p$ to a point $p'$ within the gauge orbit. The gauge generator is used to construct 
finite gauge transformations, thereby realizing active diffeomorphism-induced transformations. These gauge 
transformations define equivalence classes within ${\cal S}$, which we call gauge orbits. A gauge orbit 
represents a unique physical state\footnote{Note that this state is the whole spacetime.}, and its 
different points correspond to different coordinatizations. Certainly, one can also pass from one 
coordinatization to another by a passive diffeomorphism, but in the active view one moves the description 
from a point $p$ to a point $p'$ by changing the configuration of the fields without touching the 
coordinatization. This active view is the one that gets naturally realized in phase space with the use of 
the Poisson bracket.

\vspace{4mm}

We are ready to address the purported paradox anticipated in subsection \ref{gvd}. If we take the 
descriptors $\xi^{\mu}=N^\mu$ (which, according to (\ref{rule}), give $\epsilon^\mu=\delta^\mu_0$, the 
rigid time translation diffeomorphism), and take into account that the dynamics dictates 
$\dot N^\mu=\lambda^\mu$, then the gauge generator (\ref{thegen}) becomes exactly the Hamiltonian 
(\ref{theham}). This is the {\sl gauge versus dynamics} puzzle: the Hamiltonian is a particular case of the 
gauge generator. But, is there really any problem? Should this result really be a suprise?

To see that there is no problem at all with such a coincidence, and more than that, that this accordance 
is as it must be, just observe that in the space of on-shell field configurations the gauge generator 
moves from one point $p$ to another $p'$, whereas the Hamiltonian works within every point $p$, which 
already represents an entire spacetime. The gauge generator and the Hamiltonian can be mathematically the 
same object, but their duties are distinct. 

To illustrate, suppose we are at the point $p$ and we make an active rigid time translation $\delta t$ of all our fields - 
which in the passive view is nothing but a specific change of coordinates - we will move to another point, 
say $p'$, in the gauge orbit, in which all the fields $\Phi_{\!{}_{p'}}$ are related to the fields at $p$ 
according to $\Phi_{\!{}_{p'}} ({\bf x},t) = \Phi_{\!{}_{p}} ({\bf x},t+\delta t)$. The gauge 
transformation preserves the time coordinate and sets $\Phi_{\!{}_{p}} ({\bf x},t)\to\Phi_{\!{}_{p'}} ({\bf 
x},t)$. This transformation is indeed generated by the Dirac Hamiltonian, obtained from the gauge generator through the replacement (for all $t$), $\xi^\mu(t) \rightarrow N^\mu(t)$ and $\dot \xi^\mu(t) \rightarrow \dot N^\mu(t)$. Thus we see that the Dirac Hamiltonian does precisely what it is designed to do - it effects a global rigid translation in time on solutions of the equations of motion. But clearly this transformation does not begin to exhaust the full range of gauge transformations. In particular it does not effect a change in the functions $N^\mu$ other than global translation in time, nor the corresponding changes in the remaining phase space variables.

In connection with the analysis above, we can address also the quotienting issue mentioned in the last
paragraph in \ref{gvd}. We note again that an imprecise terminology is at the root of the problem. Let us
momentarily sit in phase space, instead of within the space of on-shell field configurations. A point in phase
space is just a configuration of the fields at a given time\footnote{This ``time'' is simply the time
coordinate for an arbitrary observer.}, and it may be intepreted as the setting of initial conditions for
the dynamics in order to build an entire spacetime.

To assert, \`a la Dirac, that there are gauge transformations
connecting points in phase space is misleading to say the least
and, should such terminology be used, it must be done under very
strict and cautious qualifications\footnote{Dirac's incomplete
analysis of gauge transformations, and the subsequent
misunderstandings generated by his work, is discussed in \cite{Pons:2004pp}}.
These transformations are not gauge symmetries in the first place,
because they do not map solutions to solutions. Thus the world
``gauge" is not used here in the usual sense. What is true is that
these transformations relate possible equivalent sets of initial
conditions, equivalent in the sense that they build, by way of the
dynamics, gauge equivalent spacetimes. It is obvious that, among
these related points in phase space, there will be points
representing future or past configurations of other points. They
all are associated with a single spacetime physics, i.e., a gauge orbit
in ${\cal S}$. Indeed, after the quotienting procedure, {\it each point in this quotient space of phase space represents precisely  a whole gauge orbit in ${\cal S}$}, and it is nonsense to try to
implement the dynamics in such space, because the dynamics takes
place at each point in the gauge orbit within {\cal S}.

In addition, one object is the spacetime physics as a whole,
represented by the gauge orbit, and another the physics locally
experienced by an observer at a given time according to his  or her
coordinatization, which we may call the ``timeslice" physics. For
an observer, this ``timeslice" physics changes along his or her own
time and when all observers agree on using the same intrinsic
coordinatization - see section (\ref{GF}) - they all describe the
same changes, which become observable. The ``frozen time" picture
claimed in subsection \ref{gvd} disappears. More details in the
appendix of \cite{Pons:2009cz}.

\subsection{Addressing the  issue of {\sl observables as constants of motion}}
%%%%%%%%%%%%%%%%%%%%%%%%%%%%%%%%%%%%%%%%%%%%%%%%%%%%%%%%%%%%%%
%%%%%%%%%%%%%%%%%%%%%%%%%%%%%%%%%%%%%%%%%%%%%%%%%%%%%%%%%%%%%%

In subsection (\ref{oct}) the claim is made that observables must have vanishing Poisson brackets with the 
the gauge generator and hence with the Dirac Hamiltonian. This is entirely correct and indisputable because,
by its very definition, observables are gauge invariant. But the next claim in subsection \ref{oct}, asserting that an observable can not depend explicitely on the coordinates is wrong. It originates from a confusion between the active and passive view of diffeomorphism invariance. In the passive view the fields are considered always the same, their mathematical description changing according to the use of different coordinatizations. Instead, in the active view, which is the view taken in phase space, a gauge transformation moves from one configuration to 
a different one without changing the coordinates. Although both views are equivalent, the coordinates 
themselves, not being variables in phase space, are gauge invariant in the active view. This explains why 
an observable may depend explicitly on the coordinates. Requiring the obervables to be independent of the 
coordinates and at the same time to have vanishing Poisson brackets with the gauge generators is to mix the 
two pictures of passive and active diffeomorphisms. It is too much of a requirement and therefore it is 
little wonder that paradoxes occur.

In conclusion, the only acceptable requirement on the observables, as a matter of their definition, is that 
they must have vanishing Poisson brackets with the the gauge generator. But they are not necessarily 
constants of motion because of their possible explicit dependences on the coordinates. Again, the ``nothing 
happens", ``frozen time" picture fades away.

\section{Gauge Fixing and scalar coordinatizations}
\label{GF}
%%%%%%%%%%%%%%%%%%%%%%%%%%%%%%%%%%%%%%%%%%%%%%%%%%%%%%%%%%%%%%
%%%%%%%%%%%%%%%%%%%%%%%%%%%%%%%%%%%%%%%%%%%%%%%%%%%%%%%%%%%%%%
We will focus on the GR case, though our considerations can be easily extended to other generally 
covariant metric theories. In a striking difference with the case of gauge theories associated with internal 
symmetries, a mandatory characteristics of the gauge fixing (GF) in generally covariant theories is that it must 
have explicit dependence on the coordinates. In particular, the explicit dependence on the time coordinate 
is a necessary condition \cite{Sundermeyer:1982gv,ps95cqg} for the GF to select a unique point in the gauge 
orbit and yet not to freeze the dynamics. Our GF constraints will be of the type 
$$
x^\mu - X^\mu(x)=0\,.
$$ 
In any gauge theory, a GF must comply with these two conditions: 
\begin{itemize}
\item
 {\sl Non-degeneracy} (or {\sl uniqueness}), that is, it must single out a unique point in the 
gauge orbit, and
\item
 {\sl Completeness}, that is, it must completely fix the dynamics. 
\end{itemize}

We examine these two issues separately for our gauge choice.

\subsection{Non-degeneracy}
%%%%%%%%%%%%%%%%%%%%%%%%%%%%%%%%%%%%%%%%%%%%%%%%%%%%%%%%%%%%%%
%%%%%%%%%%%%%%%%%%%%%%%%%%%%%%%%%%%%%%%%%%%%%%%%%%%%%%%%%%%%%%
\subsubsection{Non-degeneracy and scalar coordinatizations}
\label{physpers}
%%%%%%%%%%%%%%%%%%%%%%%%%%%%%%%%%%%%%%%%%%%%%%%%%%%%%%%%%%%%%%
%%%%%%%%%%%%%%%%%%%%%%%%%%%%%%%%%%%%%%%%%%%%%%%%%%%%%%%%%%%%%%

From the physical point of view the non-degeneracy of the GF amounts to the condition that all observers 
obtain exactly the same description for the fields - the same mathematical functions - when all them agree 
to use the intrinsic coordinatization provided by the GF. We will show that this physical condition of 
uniqueness in the description is equivalent to the condition that the fields $X^\mu$ involved in the GF must be spacetime scalars. We interpret a choice of intrinsic coordinates as a coordinate transformation from the coordinates $x^\mu$ to $X_x^\mu(x)$, where we momentarily label with a subscript $x$ the explicit functions that define the fields\footnote{We have no preconception as to whether the $X^\mu$ are fields, or field components, or local functionals of the fields. We simply call them ``fields".} $X^\mu$ as described by the observer who uses the coordinates $x^\mu$ (the $x$-observer). Suppose that instead of starting with coordinates $x^\mu$ we start instead with coordinates $y^\mu=f^\mu(x)$, for which the very same fields defining the intrinsic coordinatization are given by the functions $X_y^\mu(y)$. As seen by the $y$-observer, taking the intrinsic coordinatization amounts to the coordinate transformation from the coordinates $y^\mu$ to $X_y^\mu(y)$. Then the demand that the intrinsic coordinatization is the same for all observers is the demand that the coordinate transformation from $X_x^\mu$ to $X_y^\mu$ must be the identity transformation, i.e., 
$$
X_x^\mu(x) = X_y^\mu(y)\,, 
$$
which is the condition that the fields $X^\mu$ - which take the functional from $X_x^\mu$ for the 
$x$-observer and $X_y^\mu$ for the $y$-observer - are spacetime scalars.

\subsubsection{More on non-degeneracy}
\label{mathpers}
%%%%%%%%%%%%%%%%%%%%%%%%%%%%%%%%%%%%%%%%%%%%%%%%%%%%%%%%%%%%%%
%%%%%%%%%%%%%%%%%%%%%%%%%%%%%%%%%%%%%%%%%%%%%%%%%%%%%%%%%%%%%%

In addition to the argument above, one can easily show that there is no degeneracy, at least locally 
- that is, in a given coordinate patch - when we use scalars satisfying the completeness condition. Indeed, 
suppose that the GF constraints $x^\mu - X_x^\mu(x)=0$ are satisfied for coordinates $x^\mu$, with 
$X_x^\mu$ being, as before, the functions that express the scalar fields $X^\mu$ in the $x$ coordinates. 
Let $x\to y$ be a change of coordinates $y=f(x)$. Let us check whether the GF is satisfied for these new 
coordinates. In the $y$-coordinates the scalar fields are ${\tilde X}_y^\mu(y) = X_x^\mu(x)$. Thus, $y^\mu 
- {\tilde X}_y^\mu(y) = f^\mu(x) - X_x^\mu(x) = f^\mu(x)-x^\mu$, which can only vanish if $y^\mu=x^\mu$. 

We now give some examples of degeneracy that arises as a consequence of attempting to use intrinsic 
coordinates that do not transform as spacetime scalars. Let us start with a ``vector coordinatization". The 
GF constraints are $x^\mu - J^\mu(x)=0$, where $J^\mu(x)$ are components of a vector $J= 
J^\mu(x)\partial_{x^\mu}$. Let us proceed in full detail to show the degeneracy of this GF. Consider a 
linear transformation of the coordinate system, $x\to y$, so that $y^\mu := M^\mu_\nu x^\nu$. Then
$$
\partial_{x^\nu}=\frac{\partial y^\mu}{\partial x^\nu}\partial_{y^\mu} = M^\mu_\nu \partial_{y^\mu}\,,
$$
and therefore
$$J= J^\nu(x)\partial_{x^\nu} = J^\nu(x)M^\mu_\nu\partial_{y^\mu}
=:{\tilde J}^\mu(y) \partial_{y^\mu}\,. $$
We thus infer the obvious result that in the new coordinates $y^\mu$ the components of the vector field are
${\tilde J}^\mu(y) = M^\mu_\nu J^\nu(x)$. Notice then that
$$y^\mu - {\tilde J}^\mu(y) = M^\mu_\nu(x^\nu - J^\nu(x)) =0\,.$$
Thus if the GF constraints are satisfied for the $x^\mu$ coordinates, they will also be
satisfied for the $y^\mu$ coordinates. In consequence the GF is degenerate and hence 
it is not a good GF because it does not select a unique coordinatization. Note that this is a $GL(4,R)$ 
degeneracy, because matrices $M$ are only required to be invertible.

\vspace{4mm}

 The following  example uses a ``tensor coordinatization". We take the GF to be $x^\mu - 
T^{0\mu}(x)=0$ with $T^{0\mu}$ components of a tensor field. Considering again a linear transformation of 
coordinates $y^\mu := M^\mu_\nu x^\nu$, we now will have ${\tilde T}^{\mu\nu}(y) = M^\mu_\rho M^\nu_\sigma 
T^{\rho\sigma}(x)$, and in particular ${\tilde T}^{0\nu}(y) = M^0_\rho M^\nu_\sigma T^{\rho\sigma}(x)\,.$ 
Let us choose $M$ such that $M^0_\rho = \delta^0_\rho$. Then, 
$${\tilde T}^{0\nu}(y) =  M^\nu_\sigma T^{0\sigma}(x)\,.$$
Consequently, $y^\mu - {\tilde T}^{0\mu}(y)= M^\nu_\sigma(x^\sigma - T^{0\sigma}(x))=0$\,,
which explicitely shows the degeneracy of the GF for matrices $M$ satisfying $M^0_\rho = \delta^0_\rho$.
The  GF  $x^\mu - T^{\mu\mu}(x)=0$ can also easily be shown to be degenerate.

\vspace{4mm}

Although we do not claim to have obtained a general proof that all non-scalar coordinatizations are 
degenerate, it is clear from these examples that the procedure to show such a degeneracy is 
generalizable to many other cases, though it seems that one has to work them on a case by case basis. In 
agreement with the physical picture obtained in the previous subsection \ref{physpers}, we 
will assume from now on that only a GF made with scalar fields is acceptable.

\subsubsection{More on scalars in phase space}
\label{scalarsnolapse}
Consider general relativity coupled with matter with a non derivative coupling and with no additional gauge 
freedom. Here we prove that a scalar field in phase space - perhaps made with a local functional of the fields - can not depend on the lapse and shift. In fact, under an infinitesimal diffeomorphism $\epsilon^\mu$, 
a generic scalar $\Phi$ transforms as $\delta \Phi = \epsilon^\mu \partial_\mu \Phi$, which, using 
(\ref{rule}), can be written
\beq
\delta \Phi = \xi^0 n^\mu \partial_\mu \Phi + \xi^i \partial_i \Phi\,.
\label{delphi1}
\eeq  
On the other hand, we should be able to find the same transformation by acting with the gauge generator 
(\ref{thegen}), that is,
\beq
\delta \Phi = \{\Phi,\,G_{\xi }\} = \xi^{\mu}\{\Phi,\,{\cal H}_{\mu}+ N^{\rho} C^{\nu}_{\mu \rho} P_{\nu}\} 
+ \dot\xi^{\mu}\{\Phi,\,P_{\mu}\}\,.
\label{delphi2}
\eeq 
Due to the arbitrariness of the descriptors $\xi^{\mu}$, the equality between (\ref{delphi1}) and 
(\ref{delphi2}) implies $\{\Phi,\,P_{\mu}\}=0$, thus the scalar $\Phi$ can not functionally depend on the 
lapse and shift.

\vspace{4mm}

As a simple example consider the scalar field matter Lagrangian ${\cal L}_ 
{{}_m}=\frac{\sqrt{-g}}{2}g^{\mu\nu}\partial_\mu\phi\partial_\nu\phi$. The ADM decomposition 
\beq g_{\mu\nu}= \left(\begin{array}{c|c}- N^2+N^a N_{a}  & N_{a} \\ \hline N_{a} & \gamma_{ab}  
\end{array}\right).
\eeq
implies $\sqrt{-g}= N\sqrt{\gamma}$. The momentum $\pi = \frac{\partial {\cal L}_ {{}_m}}{\partial\dot 
\phi}$ permits us to solve for $\dot \phi$ in terms of canonical variables as 
$\dot \phi = N^a\partial_a\phi - \frac{N}{\sqrt{\gamma}}\pi$. Then, the scalar 
$g^{\mu\nu}\partial_\mu\phi\partial_\nu\phi$ becomes, in canonical variables, 
$\gamma^{ab}\partial_a\phi\partial_b\phi-\frac{\pi^2}{\gamma}$, which is free of dependences on the lapse 
and shift.

\subsection{Completeness}
%%%%%%%%%%%%%%%%%%%%%%%%%%%%%%%%%%%%%%%%%%%%%%%%%%%%%%%%%%%%%%
%%%%%%%%%%%%%%%%%%%%%%%%%%%%%%%%%%%%%%%%%%%%%%%%%%%%%%%%%%%%%%
The completeness condition is 
\beq|\{X^\mu,\, {\cal H}_\nu\}|\neq 0\,,
\label{determ}
\eeq 
which essentially indicates that the intrinsic coordinates $X^\mu$ evolve under motions generated by the 
Hamiltonian and momentum constraints. Completeness is then proven as follows. The stabilization of the 
GF constraints gives new secondary GF constraints which allow for the determination of the lapse and shift; this precisely requires that the matrix function $\{X^\mu,\, {\cal H}_\nu\}$ be invertible. Next, the 
stabilization of these secondary GF constraints determines the arbitrary functions present in the Dirac 
Hamiltonian. Thus the dynamics is totally fixed. 

\vspace{4mm}

In addition to fixing the dynamics, condition (\ref{determ}) guarantees that the fields $X^\mu$ are indeed 
a good local coordinatization, as we now show. One has, specializing the gauge generator (\ref{thegen}) to 
spacetime translations ($\epsilon^\mu = \delta^\mu_0$ for time translations and $\epsilon^\mu =  
\delta^\mu_i$ for spatial ones), 
$$\partial_\nu X^\mu = {\cal A}^\mu_{\ \rho} {\cal M}^\rho_{\ \nu}\,,
$$
with ${\cal A}$ being the invertible matrix (with discrete as well as continuous indices)
\beq
{\cal A}^\mu_{\ \rho} := \left\{ X^\mu, {\cal H}_{\rho} \right\},
\label{calA} 
\eeq
and 
$${\cal M} = \left(\begin{array}{cc} N &
\vec 0 \\
 \vec N& I\end{array} \right)\,.
$$
Since both ${\cal A}$ and ${\cal M}$ are invertible (notice that $|{\cal M}| = N > 0$), we infer that
$\partial_\nu X^\mu$ is invertible. This is the Jacobian of the transformation from coordinates $x$ (those 
used by an oberver at some arbitrary 
point $p$ in the gauge orbit) to the intrinsic coordinates defined by the scalars $X^\mu$. Thus, thanks to 
the completeness condition of the GF, that is, that ${\cal A}$ is invertible, we see that the scalars 
$X^\mu$ do indeed define a good local coordinatization. 

Note that throughout our analysis global issues concerning the use of a set of four scalar fields to provide for 
a coordinatization of the whole spacetime remain unaddressed. In general one should expect the condition 
$|\{X^\mu,\, {\cal H}_\nu\}|\neq 0$ to be satisfied locally, but perhaps not globally. This means that, in 
different coordinate patches of the manifold's chart, one may have to use different scalars to bring about 
an intrinsic coordinatization.

\subsection{Gauge fixings in GR}

Experimental tests of general relativity abound. One may ask how it is accomplished that these tests are ``gauge 
invariant", that is, that they have a true physical meaning? There are two possible answers. On one hand the result of the measurement might be expressible in a gauge invariant way, for instance when measuring the proper 
time for a particle trajectory. On the other hand, it might be that all observers agree to use the same 
scalar coordinatization (intrinsic coordinates) to express their results. This is exactly the GF procedure 
discussed in section \ref{GF}. We will see in the next section that, at least at the formal level, it is 
easy to construct observables out of the GF. Once the connection between observables and GF is understood, 
it is clear that it is enough to focus on measurements made in the framework of an intrinsic coordinate 
system. These GF procedures are in accordance with the use of special coordinate systems as is customary in 
the parametrized post-Newtonian formalism \cite{tegp,Will:2005va} and reflected in the IAU 2000 resolutions 
\cite{Soffel:2003cr} for astrometry, celestial mechanics and metrology in the relativistic framework.

We think it is worth mentioning in this context the theoretical realization of an intrinsic coordinate system by Rovelli in \cite{Rovelli:2001my}. A very similar proposal is the Galactic Positioning System, advocated among others by Misner \cite{misner}, based on four extremely stable pulsars - extremely stable meaning that its pulse count differs from its proper time by an affine transformation. For use in the neighborhood of the solar system, they should form the vertices of a topological tetrahedron - which is the configurations also considered by Rovelli - with the Earth near its center.

%%%%%%%%%%%%%%%%%%%%%%%%%%%%%%%%%%%%%%%%%%%%%%%%%%%%%%%%%%%%%%
%%%%%%%%%%%%%%%%%%%%%%%%%%%%%%%%%%%%%%%%%%%%%%%%%%%%%%%%%%%%%%
\section{Constructing observables}
%%%%%%%%%%%%%%%%%%%%%%%%%%%%%%%%%%%%%%%%%%%%%%%%%%%%%%%%%%%%%%
%%%%%%%%%%%%%%%%%%%%%%%%%%%%%%%%%%%%%%%%%%%%%%%%%%%%%%%%%%%%%%
As a first step towards the explicit form of the functional invariants we impose an intrinsic
coordinate-dependent gauge condition of the form considered in section \ref{GF}, 
$\chi^{(1) \mu} := x^\mu - X^\mu(x) = 0$, 
where the $X^\mu$ are spacetime scalar functions of the canonical fields. Our task is then to find the 
canonical transformation that moves the field variables to that location on the gauge orbit where the gauge 
conditions are satisfied.  Once we are in possession of this finite transformation we may employ it to
transform all the remaining fields. These actively transformed fields, when considered as functionals of 
the original fields, are the invariants that we seek.

We note that preservation of the gauge conditions under temporal evolution leads to  additional constraints 
$\chi^{(2) \mu}:=\delta^\mu_0 - {\cal A}^\mu_{\ \nu}  N^\nu  \approx 0$,  where  the matrix ${\cal A}^\mu_{\ 
\nu}$ has been introduced in (\ref{calA}). Following the lead of Henneaux and Teitelboim \cite{henneauxt94}, 
further exploited by Dittrich \cite{dittrich07} and Thiemann \cite{thiemann06}, we find it convenient to work 
with linear combinations $\overline{\zeta}_{(i) \nu}$ of the first class constraints $\zeta_{(1) \mu} :=  
{\cal H}_{\mu}, \zeta_{(2) \mu} :=P_\mu$  having the property that 
$\left\{ \chi^{(i) \mu}, \overline{\zeta}_{(j) \nu} \right\} \approx -\delta^i_j \delta^\mu_\nu $.  
For this purpose we need the inverse of the matrix $\left\{ \chi^{(i) \mu}, \zeta_{(j) \nu} \right\}$.
The appropriate linear combinations are therefore $\overline{P}_\mu = {\cal B}^\nu_{\ \mu}P_\nu$, where 
${\cal B}^\alpha_{\ \beta}$ is the inverse of ${\cal A}^\mu_{\ \nu}$, and $ {\overline {\cal
H}}_\nu = {\cal B}^\rho_{\ \nu}\left( {\cal H}_\rho - {\cal B}^\mu_{\ \lambda}N^\sigma 
\{A^\lambda_{\sigma},\,{\cal H}_\rho\} P_\mu \right)$. This new basis exists in general only locally.
Letting $\xi = {\cal B}^\mu_{\ \sigma}{\overline\xi}^\sigma$, the gauge generator (\ref{thegen}) 
expressed in terms of this new basis is, up to quadratic terms in the constraints that can be discarded 
``on-shell" \cite{Pons:2009cz},
\beq 
G_{\overline{\xi }}(t) ={\overline
P}_\nu\dot{\overline\xi^\nu} + {\overline {\cal
H}}_\nu{\overline\xi^\nu}. 
\eeq
The finite active gauge transformation of any dynamical field $\Phi$
takes the form 
\beq 
exp\left(  \{ - , \, G_{\overline{\xi}}\}
\right) {\Phi} = \Phi + \left\{ \Phi , G_{\overline{\xi}}\right\} +
\frac{1}{2} \left\{ \left\{\Phi, G_{\overline{\xi}}\right\},
G_{\overline{\xi}} \right\} + \cdots 
\label{phi} 
\eeq 
Since the scalars $X^\mu$ can not depend on the lapse and shift (see subsection \ref{scalarsnolapse})
the descriptors for the finite gauge transformation that transforms
$X^\mu$ to $x^\mu$ are easily seen to satisfy, on shell, 
\beq 
x^\mu = exp\left(  \{ - , \, G_{\overline{\xi}}\}
\right) {X^\mu}= X^\mu + \overline{\xi}^\mu, 
\eeq 
where $\overline{\xi}^\mu  := {\cal
A}^\nu_{\ \sigma}\xi^\sigma$ is considered a function of the spacetime
coordinates and we have made use of the fact that the Poisson
brackets ${\overline {\cal H}}_\mu$ with themselves vanish
``strongly'', i.e., they are proportional to terms at least
quadratic in the ${\overline {\cal H}}_\nu$. The descriptors
$\overline{\xi}^\mu = \chi^{(1)\mu}$ and $\dot{\overline{\xi}}^\mu =
\chi^{(2) \mu}$ may therefore be substituted into (\ref{phi}), after
the Poisson brackets have been computed, to obtain all invariant
functionals ${\cal I}_\Phi$ associated with the fields $\Phi$.

Although the following results hold for all fields \cite{Pons:2009cz} we
focus here on fields other than the lapse and shift. Then the
explicit expressions for the invariants are 
\bea 
{\cal I}_{\Phi} &\approx& \Phi + \chi^{(1) \mu}\{ \Phi , \,
 \overline{\cal H}_{\mu}\} + \frac{1}{2!}\chi^{(1) \mu}\chi^{(1) \nu}\{\{ \Phi , \,
 \overline{\cal H}_{\mu}\}, \,\overline{\cal H}_{\nu}\} + \cdots \nonumber\\
&=:&\sum_{n=0}^{\infty} \frac{1}{n!}(\chi^{(1)})^n \{ \Phi , \,
 \overline{\cal H}}\}_{(n)\,, \label{expansion}
\eea 
where $\approx$ is the symbol for Dirac's weak equality, that
is, an equality which holds ``on-shell". 

With different notation, this expression (\ref{expansion}) appeared in the literature in \cite{thiemann06} 
as his equation (2.8) and in \cite{dittrich07}  as her equation (5.23). Here we have arrived at
(\ref{expansion}) by a symmetry-inspired procedure, as the effect of the finite gauge transformation that 
sends any point in the space of on shell field configurations to the specific point satisfying the GF 
condition. This specific gauge transformation is determined once the set of
scalar fields associated with the gauge fixing has been selected. An advantage of the present formulation 
is that one can send all the fields from the coordinate description of any observer to the description 
given at the point of the gauge orbit where the gauge conditions are satisfied, and this includes the lapse 
and shift. 

The time rate of change of these invariant functionals satisfies \cite{Pons:2009cz}
\beq
\frac{d}{d\,t} {\cal I}_{\Phi}  \approx {\cal I}_{\{ \Phi , \, \overline{\cal H}_0\}}\,.
\label{timeder}
\eeq  
Recognizing that the invariant fields are simply the fields expressed in intrinsic coordinates, that is, 
the fields as seen by the observer sitting at the point in the gauge orbit where the GF constraints are 
satisfied, we note that 
${\cal I}_{\{ \Phi ,\,\overline{\cal H}_0\}}= {\{ \Phi , \,
\overline{\cal H}_0\}}_{p_{{}_{\!G}}} = \{ \Phi , \,N^\mu {\cal H}_\mu\}_{p_{{}_{\!G}}}$. 
Therefore the invariants satisfy the equations of motion of the gauge fixed fields. We note also that
repeated time derivatives of the invariant fields yield the simple expression 
$\frac{\partial^n}{\partial\,t^n} {\cal I}_{\Phi}\approx{\cal I}_{\{ \Phi , \,
\overline{\cal H}_0\}_{(n)}}$, and that we may therefore express the invariants as follows as a
Taylor series in $t$: 
\beq 
{\cal I}_{\Phi}\approx\sum_{n=0}^{\infty} \frac{t^n}{n!}\,  {\cal I}_{\{ \Phi ,
\,\overline{\cal H}_0\}_{(n)}}{}_{\big|_{t= 0}}\,. \label{evolv}
\eeq 
This expression displays in a striking manner the notion of {\it evolving constant of the motion} 
introduced originally by Rovelli \cite{rovelli90}. We stress however, that the evolution we obtain here is 
nothing other than the evolution determined by Einstein's equations in the gauge-fixed coordinate system. 
In addition, the potentially infinite set of invariants ${\cal I}_{\{ \Phi , \,\overline{\cal 
H}_0\}_{(n)}}$ will correspond to Noether symmetries \cite{Batlle:1987ek,Pons:2009cz}. In generic general 
relativity the conserved quantities must necessarily be spatially non-local, as pointed out by 
Torre \cite{torre93}.

Notice that the coefficients of the expansion (\ref{evolv}) in the time parameter are constants of motion, 
because they are invariants without explicit time dependence and, since they have vanishing Poisson 
brackets with the Hamiltonians, they have no implicit time dependence either. In fact, fixing the explicit 
time parameter at any arbitrary value, the invariants ${\cal I}_{\Phi}$ become constants of motion.

\vspace{4mm}

To close this section we mention two relevant results from \cite{Pons:2009cz}. 

Clearly, the map $\Phi\to {\cal I}_{\Phi}$ is not a canonical transformation because it sends all the gauge equivalent configurations
to the same configuration, the one satisfying the gauge fixing conditions. It can nevertheless be understood as a limit of a one-parameter family of canonical maps. Let us consider the functionals, obtained from canonical maps,
${\cal K}_{\Phi}^{(\Lambda)} = exp\left(  \{ - ,\Lambda\,\chi^{(1) \nu} \overline{\cal H}_{\nu}\} \right)\Phi$. One can prove that, on shell, \cite{Pons:2009cz}
\beq {\cal K}_{\Phi}^{(\Lambda)}
\approx \sum_{n=0}^{\infty}\frac{1}{n!}(1-e^{-\Lambda})^n(\chi^{(1)})^n \{ \Phi , \,
 \overline{\cal H}\}_{(n)}\,,
\label{theK}
\eeq
from which it follows that 
$$\lim_{\Lambda\rightarrow\infty}{\cal K}_{\Phi}^{(\Lambda)} \approx {\cal I}_{\Phi}\,.
$$

\vspace{4mm}

The second result is the observation that the Poisson brackets of the invariants are simply the invariants associated with the Dirac brackets of the fields. This result, already obtained in \cite{thiemann06} for fields other than lapse and shift, holds in fact for for all fields,
\beq 
\{{\cal I}_{\Phi^A},\,{\cal I}_{\Phi^B}\} \approx {\cal I}_{\{\Phi^A,\,\Phi^B\}^*}\,. 
\eeq

\section{An example: observables for a massive particle in an AdS background}
%%%%%%%%%%%%%%%%%%%%%%%%%%%%%%%%%%%%%%%%%%%%%%%%%%%%%%%%%%%%%%
%%%%%%%%%%%%%%%%%%%%%%%%%%%%%%%%%%%%%%%%%%%%%%%%%%%%%%%%%%%%%%

Consider the massive free particle in AdS spacetime. In Poincar\'e coordinates (actually Poincar\'e coordinates use $u=\frac{1}{r}$), 
$$
ds^2= g_{\mu\nu}dq^{\mu}dq^{\nu}= \frac{R^2}{r^2}(\eta_{\mu\nu}dq^{\mu}dq^{\nu})= 
\frac{R^2}{r^2}(\eta_{\alpha\beta}dq^{\alpha}dq^{\beta}+ dr^2)
$$
with $\mu=(\alpha, n-1),\quad r= q^{n-1},\quad \alpha=(0,\ldots,n-2),\quad 
\eta_{\alpha\beta}=(-,+,\ldots,+)$\,.
These coordinates only describe a patch of the whole AdS spacetime. 
Now $r\geq 0$ and $r=0$ is the boundary.

The Lagrangian for the massive particle in the AdS background is
$$ 
L = \frac{1}{2\,N}g_{\mu\nu}\dot q^\mu\dot q^\nu - \frac{1}{2} m^2 N\,,
$$ 
with $N$ an auxilary variable. The Dirac Hamiltonian is
$$ 
H_{\!D} = \frac{1}{2} N (g^{\mu\nu} p_\mu p_\nu +  m^2) + \lambda \pi,
$$ 
where $\pi$, the momentum canonically conjugate to $N$, is the primary constraint, and $\lambda$ an 
arbitrary function of time. There is a secondary constraint, namely ${\cal H} = \frac{1}{2} (g^{\mu\nu} 
p_\mu p_\nu + m^2)$. The gauge generator has the form $G = \xi\, {\cal H} + \dot \xi \pi$. We choose as a gauge fixing constraint $\chi = t-q^0$. Next, following the usual procedure, we define 
$\displaystyle A:= \{q^0,\,{\cal H}\} = \frac{r^2}{R^2}p^0$ (for our purposes, here and henceforth, indices 
are raised and lowered exclusively with $\eta^{\mu\nu},\ \eta_{\mu\nu}$.), and 
$$
\overline{\cal H} := \frac{1}{A}{\cal H} =\frac{1}{2\,p^0}(\eta^{\mu\nu} p_\mu p_\nu + 
\frac{R^2m^2}{r^2})\,.
$$ 
Now we are ready to compute the invariants associated with the coordinates and the momenta. Note that we do 
not write the implicit time dependence in the variables, which is the same as if we were working just in 
phase space instead of working in the space of trajectories - i.e., field configurations. 

Most of the series expansions are trivial. So we get (for $\alpha=(0,i)$)
\bea 
{\cal I}_{p_\alpha} &=& p_\alpha\,,\nonumber \\
{\cal I}_{q^i} &=& q^i + \chi \{q^i,\,\overline{\cal H}\} = q^i + (t-q^0) \frac{p^i}{p^0} \nonumber \\
&=&  (q^i-\frac{p^i}{p^0}q^0) + \frac{p^i}{p^0} t  \,. 
\eea
Thus we extract from ${\cal I}_{p_\alpha}$ the constants of motion (CM) $p_\alpha$. On the other hand, 
the expansion in the $t$ parameter for ${\cal I}_{q^i}$ gives the CM $q^i-\frac{p^i}{p^0}q^0$ and 
$\frac{p^i}{p^0}$. Using the CM $p_\alpha$ we can identify the CM $C^{0\alpha}:= q^0 p^\alpha-p^0 q^\alpha$.
To determine the rest of the basic invariants it is convenient to compute ${\cal I}_{r^2}$ instead of 
${\cal I}_{r}$ (In fact ${\cal I}_{r^2}= ({\cal I}_{r})^2$). We get the finite series
$$
{\cal I}_{r^2}= r^2 + 2\,\frac{t-q^0}{p^0}\, r\, p_r + (\frac{t-q^0}{p^0})^2(p_r^2 + 
\frac{R^2\,m^2}{r^2})\,,
$$
out of which we identify, as coefficients of the $t$-expansion, the constants of motion
\bea C_{{}_{(1)}}&:=& p_r^2 + \frac{R^2\,m^2}{r^2}\nonumber \\
C_{{}_{(2)}}&:=& r\,p_r - \frac{q^0}{p^0}(p_r^2 + \frac{R^2\,m^2}{r^2})\nonumber \\
C_{{}_{(3)}}&:=& r^2 -2\,\frac{q^0}{p^0}\,r\,p_r + (\frac{q^0}{p^0})^2 (p_r^2 + \frac{R^2\,m^2}{r^2})\,.
\eea
Note that, due to the Hamiltonian constraint, $C_{{}_{(1)}}\simeq -\eta^{\alpha\beta} p_\alpha p_\beta$, 
that is, 
$C_{{}_{(1)}}$ is already a combination of other CM. On the other hand, 
$\displaystyle \frac{1}{2\,p^0} C_{{}_{(1)}}\,C_{{}_{(3)}} = C_{{}_{(2)}}^2 + R^2\,m^2$; 
therefore we only get $C_{{}_{(2)}}$ as a new constant of motion. Note also that, using the Hamiltonian 
constraint,
\bea 
C_{{}_{(2)}} &\simeq& r\,p_r + \frac{q^0}{p^0}(\eta^{\alpha\beta} p_\alpha p_\beta) =
r\,p_r + \frac{1}{p^0} p_\alpha (q^0 p^\alpha) \nonumber \\ 
&=& r\,p_r + \frac{1}{p^0} p_\alpha (C^{0\alpha}+ p^0 q^\alpha) = \frac{p_\alpha}{p^0} C^{0\alpha} + p_\mu 
q^\mu\,,
\eea
thus we have obtained the CM $S:= p_\mu q^\mu$, which is the generator of the scale transformations.

Finally, due to the Hamiltonian constraint, the computation of ${\cal I}_{p_r}$, or better ${\cal 
I}_{p_r^2}$, does not yield any new CM's. Nothing new appears either for ${\cal I}_{N}$.

\vspace{4mm}

Summarizing, we have obtained $n-1$ momenta, $p_\alpha$, $n-2$ boost generators, $C^{0\, i}$, and the 
generator $S$ of scale transformations. These constants of motion are 
functionally independent and they number $2n-2$. Fixing the values of these CM - such that they satisfy the 
Hamitonian constraint\footnote{In fact we only need to care that $\eta^{\mu\nu} p_\mu p_\nu < 0$ because 
the mass $m$ is arbitrary and it does not appear in the CM mentioned above.} - does not determine 
completely the trajectory of the particle. For this we need another independent CM. Actually, it is 
provided by one of the generators of special conformal transformations. We may take, for instance:
$C^0 := 2 q^0 S - p^0 \eta_{\mu\nu} q^\mu q^\nu$. Once we are in possession of $2n-1$ independent CM, since 
they fix a trajectory for the particle in phase space - again, chosing values compatible with the 
Hamiltonian constraint - it is clear that all other CM must be functionally dependent of this set of $2n-1$ 
independent CM. For instance, the generators of boosts and rotations, $C^{\alpha\beta}:= q^\alpha p^\beta-p^\alpha q^\beta$, can be written as $C^{\alpha\beta}= 
\frac{1}{p^0}(p^\alpha C^{0\,\beta} -p^\beta C^{0\,\alpha} )$, and the 
generators of special conformal transformations, $C^\alpha := 2 q^\alpha S - p^\alpha \eta_{\mu\nu} q^\mu q^\nu$, are related to other CM by 
$$
C^\alpha - \frac{1}{p^0}(p^\alpha C^0- 2 C^{0\,\alpha} S )= 0\,.
$$
One can also verify the relationship among these CM
$$
p^\alpha C^\beta -p^\beta C^\alpha + 2 C^{\alpha\beta} S =0\,.
$$
This discussion for the AdS background is in contrast with the study of the massive free particle in a 
Minkowski background, undertaken in \cite{Pons:2009cz}. In this case, with the same procedure outlined above, one 
obtains as CM extracted from the observables, $n$ generators of translations and $n-1$ generators of 
boosts, totaling $2n-1$ independent CM. With the usual restriction of imposing compatiblility with the 
Hamiltonian constraint, they will determine a trajectory in phase space. All other CM must then be 
functionally dependent of the CM derived from the observables. 

\vspace{4mm}

\subsection{Constants of motion and Killing vectors}

It is easy to show that the CM that happen to be linear in the momenta are associated with the Killing 
symmetries of the background. Indeed, consider $K^\mu(q)p_\mu$ a candidate to be a CM. Its Poisson bracket 
with the Hamiltonian constraint is 
$$\{K^\rho(q)p_\rho,\,{\cal H}\}=\{K^\rho(q)p_\rho,\,\frac{1}{2} (g^{\mu\nu} p_\mu p_\nu + m^2)\} = -({\cal 
L}_{{}_K} g^{\mu\nu}) p_\mu p_\nu
$$
where ${\cal L}_{{}_K} g^{\mu\nu} $ is the Lie derivative of the metric under the vector field 
$K^\rho(q)\partial_\rho$. Thus
$$\{K^\rho(q)p_\rho,\,{\cal H}\}= 0\quad \Longleftrightarrow\quad  {\cal L}_{{}_K} g^{\mu\nu}=0\,.
$$

Applied to our AdS case, we see that our observables account for the Killing vectors associated with 
translations ($\alpha$ directions), boost ($i$ directions), rotations ($i,j$ directions), and the scale 
transformation. The only missing Killing vectors are those associated with the special conformal 
transformations, which have not been obtained within our method. It is perhaps possible that another choice 
for the gauge fixing could account for these CM, but we do not know for sure. In fact, the choice $\chi = 
t-r$ as the gauge fixing yields the observable
$$ {\cal I}_{p_r^2} = p_r^2 + \frac{R^2\,m^2}{r^2}- \frac{R^2\,m^2}{t^2}\,,
$$
which does not introduce new CM.

\section{Conclusion: a case closed}
%%%%%%%%%%%%%%%%%%%%%%%%%%%%%%%%%%%%%%%%%%%%%%%%%%%%%%%%%%%%%%
%%%%%%%%%%%%%%%%%%%%%%%%%%%%%%%%%%%%%%%%%%%%%%%%%%%%%%%%%%%%%%
We have shown that it is possible to construct, albeit in a formal way, observables in general relativity by employing a gauge fixing using a scalar coordinatization. In this way we obtain a new understanding as to why finding observables in generally covariant theories is such a difficult mathematical task. But once these two points  - the existence and the difficulty of construction of observables - have been made, a new vision emerges: that constructing these observables through the use of active diffeomorphism-induced symmetry transformations - which are valid for every observer with his/her own coordinatization - is not the most efficient procedure. Indeed, once we have proven that observables can be built for any observer, we can gladly dispose of this construction and just take the passive view of diffeomorphism invariance. We simply instruct each observer, having constructed his or her phase space solutions, to transform them to the intrinsic coordinate system! We have indeed proven that the final result is coincident with the active construction. Of course one benefit of the active construction is that it can be used to prove that the resulting invariants have the property that their Poisson brackets are precisely the invariants associated with the Dirac brackets
of the fields. This construction procedure was in fact undertaken in \cite{ps04} for the case of the relativistic free particle and in \cite{shs07} for a homogeneous anisotropic cosmological model with a massless scalar field source.

In the gauge orbit picture within the space of on shell field configurations, this agreement has the effect of choosing the gauge fixed description, that is, of moving to the only point in the gauge orbit where the 
GF constraints are satisfied. Indeed, fixing the gauge by selecting a coordinate system is the standard procedure used for the experimental tests of general relativity \cite{tegp,Will:2005va}.

Thus here is the guiding principle: {\it let everyone adopt the same instrinsic coordinates}. Once this instruction is implemented all geometric objects become observable! All observers attain the same description regardless of the coordinate system with which they begin their construction. In other words, the final description is invariant under alterations in this initial arbitrary coordinate choice.

In this sense we believe that our analysis of the connection between the gauge fixing procedures and the construction of observables closes the case of the problem of observables in general relativity. The accompanying ``mysteries" of ``frozen time" and ``nothing happens" have faded away. Indeed, the case is closed.

\end{document}